\documentclass{ws-procs961x669}            
\begin{document}
\title{The passage of time and top-down causation}

\author{Barbara Drossel $^*$ }

\address{Institute of Condensed Matter Physics, Technical University of Darmstadt,\\
Hochschulstr. 6, 64289 Darmstadt, Germany\\
$^*$E-mail: drossel@fkp.tu-darmstadt.de}

\begin{abstract}
  It is often claimed that the fundamental laws of physics are
  deterministic and time-symmetric and that therefore our experience
  of the passage of time is an illusion. This talk will critically
  discuss these claims and show that they are based on the
  misconception that the laws of physics are an exact and complete
  description of nature. I will argue that all supposedly fundamental
  deterministic and time-symmetric laws have their limitations and are
  supplemented by stochastic and irreversible elements. In fact, a
  deterministic description of a system is valid only as long as
  interactions with the rest of the world can be ignored. The most
  famous example is the quantum measurement process that occurs when a
  quantum system interacts with a macroscopic environment such as a
  measurement apparatus. This environment determines in a top-down way
  the possible outcomes of the measure- ment and their
  probabilities. I will argue that more generally the possible events
  that can occur in a system and their probabilities are the result of
  top-down influences from the wider context. In this way the
  microscopic level of a system is causally open to influences from
  the macroscopic environment. In conclusion, indeterminism and
  irreversibility are the result of a system being embedded in a wider
  context.
\end{abstract}


\bodymatter

\section{Introduction}
The success of physics at explaining, calculating and controling processes in nature has led to the widespread belief that the equations of physics describe accurately how the state of a system changes with time under the influence of the various physical forces. Even more, many physicists think that the properties of the fundamental equations of physics are also properties of nature. Since these equations are deterministic, it is concluded that nature also is deterministic, with the time development being completely determined by the initial state and the fundamental laws. Furthermore, these laws are invariant under changing the direction of time. Based on this, many physicists think that the perceived direction of time is merely a consequence of very special initial conditions of the universe, and that all processes in nature could in principle also run backwards in time. Einstein is the most prominent scientist who considered the irreversible passage of time as an illusion since the present contains already the future, as it is fixed by the deterministic laws of physics and the initial conditions.

The implications of this understanding of physics are immense. It follows in particular that our experience of being agents that can act in the ``now'' and thus bring about future developments is also an illusion. And if everything is fully determined by the laws of physics, this contradicts our experience that our decisions are driven by non-physical causes such as values and logic and purpose. 

While many philosophers of science buy the bold claims made by physicists, others are more skeptical. Indeed, there are many reasons to be skeptical. Physics as a science is always an approximate and incomplete description of nature and not a direct image of what goes on in nature. And many physicists not working with supposedly 'fundamental' physics such as quantum physics, but in condensed matter physics or soft matter physics, are aware of these limitations of the basic physical theories at describing nature. Instead of deriving the properties of their systems starting from the many-particle Schr\"odinger equation (which would be the fundamental equation for an object that consists of a huge number of atoms), they write down simpler models and effective theories that capture more directly the phenomenon to be described. These theories are often even in logical contradiction with the supposed fundamental theory, as explained in the following quote by Nobel laureate Anthony Leggett \citep{leggett1992nature}:

\begin{quote} No significant advance in the theory of matter in bulk
has ever come about through derivation from microscopic principles. [...] I would confidently argue further that it is in principle
and forever impossible to carry out such a derivation. [...] The so-called derivations of the results of solid state
physics from microscopic principles alone are almost all bogus, if
'derivation' is meant to have anything like its usual sense. Consider as
elementary a principle as Ohm's law. As far as I know, no-one has ever
come even remotely within reach of deriving Ohm's law from microscopic
principles without a whole host of auxiliary assumptions ('physical
approximations'), which one almost certainly would not have thought of
making unless one knew in advance the result one wanted to get, (and
some of which may be regarded as essentially begging the question). This
situation is fairly typical: once you have reason to believe that a certain
kind of model or theory will actually work at the macroscopic or intermediate
level, then it is sometimes possible to show that you can 'derive'
it from microscopic theory, in the sense that you may be able to find the
auxiliary assumptions or approximations you have to make to lead to the
result you want. But you can practically never justify these auxiliary
assumptions, and the whole process is highly dangerous anyway: very often
you find that what you thought you had 'proved' comes unstuck
experimentally (for instance, you 'prove' Ohm's law quite generally only
to discover that superconductors don't obey it) and when you go back to
your proof you discover as often as not that you had implicitly slipped in
an assumption that begs the whole question. [...] 
I claim then that the important advances in macroscopic physics come
essentially in the construction of models at an intermediate or macroscopic
level, and that these are logically (and psychologically) independent of
microscopic physics. Examples of the kind of models I have in mind which
may be familiar to some readers include the Debye model of a crystalline
solid, the idea of a quasiparticle, the Ising or Heisenberg picture of a
magnetic material, the two-fluid model of liquid helium, London's
approach to superconductivity .... In some cases these models may be
plausibly represented as 'based on' microscopic physics, in the sense that
they can be described as making assumptions about microscopic entities
(e.g. 'the atoms are arranged in a regular lattice'), but in other cases (such
as the two-fluid model) they are independent even in this sense. What all
have in common is that they can serve as some kind of concrete picture,
or metaphor, which can guide our thinking about the subject in question.
And they guide it in their own right, and not because they are a sort of
crude shorthand for some underlying mathematics derived from 'basic
principles.' 
\end{quote}

The goal of this paper is to show that the claim that nature is at its most fundamental level reversible and deterministic is wrong.  First, I want to explain that the time-reversible and deterministic nature of the supposedly fundamental physical equations are in fact idealizations that hold only under very specific conditions. In particular, the studied objects must be carefully isolated from interacting with the rest of the world. By taking a closer look at the areas of physics that are supposedly based on these fundamental equations, I will point out instances where indeterministic and irreversible additional elements are added to the theory when needed. This shows that the programme of deriving everything \textit{only} from the fundamental equations is never realized in practice.
I will furthermore argue that indeterminism and irreversibility are closely related, as the first implies the second.

The central part of my paper will focus on the influence of the environment
and on top-down effects. By looking at what happens in a quantum measurement,
we will see that the environment determines the possible events and their
probabilities. This will lead us to a more general discussion of how
stochasticity and irreversibility are the result of top-down effects from the
context. A consequence of this is a contextually emergent view of nature,
where not everything is controlled bottom-up by microphysics but there are
also top-down influences from the surrounding context - even from without physics.
This prepares the conclusion that while physics underlies everything that happens in nature, it does not determine everything.

\section{Irreversibility and indeterminism in the 'fundamental' theories of physics}
The following considerations show that no theory that was once or is now thought to be fundamental can do without adding irreversible elements.

\subsection{Classical mechanics}

Classical mechanics is based on Newton's law
\begin{equation}
m \frac{d^2 \vec x}{dt^2} = \vec F(\vec x)\, .
\end{equation}
This law is deterministic as the position $\vec x$ and velocity $\vec v = d\vec x /dt$ at a given moment in time determine the future time evolution of $\vec x$ and $\vec v$. The law is invariant under time reversal as a change of the sign of $t$ does not change the law. Consequently, when a time evolution of $\vec v$ and $\vec x$ has occurred from time $t=0$ to $t=t_f$, the reverse time evolution takes place when starting at time $t_f$ and inverting the velocity $\vec v$. When applying Newton's law to many particles that interact via mutual forces, the same conclusions are obtained. Due to its impressive success at unifying Kepler's laws of planetary motion and Galileis laws for falling bodies, Newton's theory was considered for a long time an exact and comprehensive description of the physical world, but there have always been cautious voices pointing out its limitations, see \cite{van2021physics,del2021indeterminism}. Only with the advent of the theory of relativity and quantum mechanics did it become clear to everybody that this view is wrong. Sometimes I wonder why physicists still make this mistake to believe that their most recent theories are exact and universal....

Back to classical mechanics: Even before the mentioned developments in the
20th century, it was clear that classical mechanics cannot do without irreversible elements: When comparing the equations to reality, one needs to include friction. For instance, a pendulum that swings freely under the influence of gravity will eventually come to rest as friction reduces the swinging amplitude with time. When including friction, Newton's law becomes
\begin{equation}
m \frac{d^2 \vec x}{dt^2} = \vec F(\vec x) - \eta \frac{d\vec x}{dt}
\end{equation}
with a friction term proportional to the velocity, the strength of which is determined by the friction coefficient $\eta$. Even the apparently so perpetual motion of the planets on their orbits is slowed down by friction due to tidal forces. 

Classical mechanics also knows stochasticity. As explained so well by Gisin \cite{gisin2019indeterminism,gisin2020real}, determinism is valid only if position and velocity have infinite precision. But physics is always limited to a finite number of bits when measuring and calculating the motion of objects. For chaotic systems, the time evolution of which depends extremely sensitively on the initial values, this means in practice that prediction of the future time evolution becomes impossible beyond a certain time horizon. There are good reasons to conclude that this indeterminism is not just a matter of our limited abilities, but an inherent feature of nature, see also \citep{drossel2015relation,del2019physics,gisin2020real}. 

If we accept that chaotic systems are inherently stochastic, the irreversible nature of friction is coupled to stochasticity on the atomic level of description: Viewed microscopically, the slowing down of the pendulum is due to collisions of the pendulum with the randomly moving molecules in the air. However, the motion of the atoms is not deterministic since it is chaotic when described by Newton's laws. 

\subsection{Classical electrodynamics}
Classical electrodynamics is governed by Maxwell's four equations,
\begin{eqnarray}
    \label{eq:Mv1}
    \nabla \cdot \vec E &=& 4 \pi \varrho
    \\
    \label{eq:Mv2}
    \nabla \times \vec E &=& - \frac{1}{c} \frac{\partial \vec B}{\partial t}
    \\
    \label{eq:Mv3}
    \nabla\cdot \vec B &=& 0
    \\
    \label{eq:Mv4}
    \nabla \times \vec B &=& \frac{4 \pi}{c} \vec j + \frac{1}{c} 
    \frac{\partial \vec E}{\partial t}
\end{eqnarray}
These equations are again deterministic as an initial configuration of the electric and magnetic  fields $\vec E$ and $\vec B$ and of the charge and current density distributions $\varrho$ and $\vec j$ determine the future evolution of the fields, and using Newton's equation for the motion of the charges under the influence of the fields they also determine the time evolution of the charge and current densities.

The equations are also time reversible: When reversing the direction of time and that of currents and magnetic fields, Maxwell's equations remain unchanged. So for each time evolution there is an equally realistic reversed time evolution (as nature knows of no preferred oriention for currents or magnetic fields).

However, there is an important subfield of electrodynamics where the time reversal invariance is broken: This is the emission of radiation from accelerated charges. Using the electrostatic potential $\phi$ and the magnetic vector potential $\vec A$ instead of $\vec E$ and $\vec B$, the emitted radiation takes the form
\begin{eqnarray}
    \phi(\vec r, t) &=&
    \int d ^3 r' 
    \frac{\varrho \left( \vec r\:', t - \frac{|\vec r - \vec r\:'|}{c} \right)}
        {|\vec r - \vec r\:'|}\, ,\label{phiret}
    \\
    \vec A (\vec r, t) &=&
    \frac{1}{c}
    \int d ^3 r'
    \frac{\vec j \left( \vec r\:', t - \frac{|\vec r - \vec r\:'|}{c} \right)}
        {|\vec r - \vec r\:'|}\, .\label{aret}
\end{eqnarray}
This means that the potentials (and consequently the fields) at a position $\vec x$ at time $t$ are determined by the motion of the charges at earlier times $ t - \frac{|\vec r - \vec r\:'|}{c}$, which is simply the time electromagnetic radiation takes to propagate with the velocity of light from the point of origination $\vec r\:'$ to the point of measurement  $\vec r$. All electromagnetic fields thus have localized sources in the temporal past \cite{weinstein2011electromagnetism}.

When explaining the reasons why emission of radiation is described by the retarded solutions, one has to resort to microscopic stochasticity: Radiation emitted by localized sources is absorbed by walls etc. The reverse process would be  that different walls conspire to emit radiation that converges from all directions to a localised sink where it is completely absorbed and turned into motion of electrical charges. However, this 'conspiracy' is argued to be impossible as the atoms of the walls perform random thermal motion and therefore emit incoherent thermal radiation and not radiation that is correlated over large spatial scales. The thermal motion of the wall atoms is considered to be stochastic. 

There is a second way how stochasticity enters electrodynamics: When trying to describe what happens at the atomic scale
when light is emitted, one needs to resort to a quantum version of
electrodynamics, which is quantum electrodynamics and describes the emission
of light in terms of the stochastic emission of energy quanta (each containing
an energy $\hbar \omega$ with $\omega$ being the frequency) from the source.
And this brings us to the next section, which deals with quantum mechanics.

\subsection{Quantum mechanics}

The basic equation of nonrelativistic quantum mechanics for one particle is the Schr\"odinger equation
\begin{equation}
i \hbar \frac{\partial}{\partial t}\Psi(\vec x,t) = \hat H \Psi(\vec x)\label{fullschroedinger}
\end{equation}
with the Hamilton operator
\begin{equation}
\hat H= \frac{\hat p^2}{2m} + V(\vec x) \, .\label{fullhamiltonian}
\end{equation}
Here, the first term is the kinetic energy, and the second the external potential. Given the initial state $\psi(\vec x,0)$, the state at other times $t$  are given by the relation
\begin{equation}
\Psi(\vec x,t) = e^{-i\hat H t/\hbar}\, \Psi(\vec x,0) \label{timeevolution}
\end{equation}
if the potential $V(\vec x)$ is not  explicitly time dependent.

This equation is deterministic. This means that the initial state, combined with the Hamiltonian determines the future time evolution. It is also invariant under time reversal:  The time-reversed Schr\"odinger equation is solved by the complex conjugate wave function $\Psi^*$, and this does not affect the observables as they are calculated from expressions that contain  products of $\Psi$ and $\Psi^*$. All this holds also for the many-perticle version of the Schr\"odinger equation. 

However, quantum mechanics is incomplete without a rule for how to calculate the outcome of a measurement. And this rule says that out of all possible measurement outcomes of an observable, one of them (let us call the corresponding state $\phi_n$) is chosen stochastically with a probability that is given by $|\langle \phi_n | \Psi\rangle |^2$. This process is irreversible, as the reverse process (that the measurement apparatus or photographic plate returns to the pre-measurement state and emits the particle that has been absorbed during measurement) is not observed in nature. In this way, quantum mechanics includes from the onset irreversibility and stochasticity. There are a number of interpretations of quantum mechanics that attempt to explain away this dichotomy between the Schr\"odinger equation and the measurement process by describing also all the atoms of the measurment device by quantum mechanics, usually invoquing decoherence in one way or another \cite{zurek2003decoherence}; however, to many people these interpretations remain unsatisfactory as they cannot really explain why in a single run of an experiment only one of the possible outcomes is observed \citep{schlosshauer2005decoherence,drossel2018contextual}. In one way or another, stochasticity kreeps into any interpretation. It does so even in Bohmian mechanics, where all stochasticity of the future time evolution is contained in the random features of the initial state. 

\subsection{Quantum field theory}

The theory that is often considered the most fundamental one is quantum field theory, which is a relativistic theory for many particles and has several building blocks that take into account the different types of interactions and particles. It is a combination of the Glashow-Salam-Weinberg model for the electroweak interaction and of quantum chromodynamics for the strong interaction. Just as for nonrelativistic quantum mechanics, this theory includes two parts. Unitary time evolution according to a Hamilton operator is only applied  between preparation and measurement of particles. The measurement outcome (e.g. of a scattering experiment or  particle collision experiment) is again described by a probability that is calculated in a  similar way as above. This means that everything written above for nonrelativistic quantum mechanics applies also to quantum field theory.

\subsection{Thermodynamics}

Thermodynamics is considered fundamental only by some scientists. But these scientists argue that thermodynamics might be even more fundamental than the other fields of physics. The most famous quotation in this direction is by Sir Arthur Eddington (From his book \textit{The Nature of the Physical World} \citep{eddington2019nature}):
\begin{quote}
  The law that entropy always increases holds, I think, the supreme position among the laws of Nature. If someone points out to you that your pet theory of the universe is in disagreement with Maxwell's equations - then so much the worse for Maxwell's equations. If it is found to be contradicted by observation - well, these experimentalists do bungle things sometimes. But if your theory is found to be against the Second Law of Thermodynamics I can give you no hope; there is nothing for it but to collapse in deepest humiliation.
  \end{quote}

Einstein had a similary high opinion of thermodynamics: 
 \begin{quote} A law is more impressive the greater the simplicity of its premises, the more different are the kinds of things it relates, and the more extended its range of applicability. [...] It is the only physical theory of universal content, which I am convinced, that within the framework of applicability of its basic concepts will never be overthrown.\end{quote}

Now, the second law of thermodynamics is about the irreversibility of nature: All processes run in the direction in which the entropy of the universe increases. The relations of thermodynamics can be obtained also from a microscopic description, which is that of statistical physics. But statistical physics is not deterministic, as it is based on probabilities for the different possible microscopic states a system can take. This means that we have again a link between irreversibility and stochasticity in this field of physics.

 \section{A reversible, deterministic theory cannot give irreversibility}

It is often argued that a 'fundamental' microscropic theory that is deterministic and reversible, such as classical mechanics or quantum mechanics, can give rise to irreversibility. Even textbooks on statistical mechanics often make this claim and use arguments based on coarse-graining. In this section, I will demonstrate that the 'derivations' of irreversibility from  a microscopic reversible and deterministic theory all employ additional assumptions that are not part of the theory. They all invoque stochasticity or randomness in one way or another and are therefore not truly deterministic. As soon as stochasticity is employed, irreversibility follows naturally. The first subsection will argue that stochasticity leads to irreversibilty, and the second subsection will show how stochasticity is smuggled into the derivations of the second laws of thermodynamics from classical mechanics.

\subsection{Stochasticity gives irreversibility}

In the previous section, we have seen that irreversible processes are always coupled to stochasticity when considered on the atomic level, for instance when friction is described in terms of atomic collisions or when the approach to thermodynamic equilibrium is described by stochastic transitions between different microscopic states. More general considerations suggest that irreversibility and stochasticity are ideed  intimately connected: Progress of time is perceived via changes. These changes occur in the form of events. Whenever one of a set of possible events occurs, the open future becomes the definite past. Thus, events are both irreversible and stochastic. There are several theories that are based on such events constituting the fabric of spacetime itself on the Planck scale  \citep{cortes2014quantum,sorkin1990spacetime}. But we need not know the most fundamental level in order to see that any stochastic event is also irreversible. This is because once the event has occured the state prior to this event cannot be retrodicted. For a quantum measurement, e.g., the outcome does not allow to reconstruct the incoming wave function. A stochastic event itself is a step by which an indeterminate future gives way to a definite outcome. Against this, it is sometimes argued that probabilities can also be applied backwards for retrodiction. Yes, but this is qualitatively different: probabilities used in retrodiction are either Bayesian probabilities that are based on incomplete knowledge of the past, or they are frequencies in an ensemble of events. In principle, the past that precedes a given event is fixed in either case. In contrast, the future is not fixed if stochasticity is real and not merely apparent. 

\subsection{The hidden assumption in so-called 'derivations' of the Second Law from classical mechanics}

In the following, I will use classical mechanics to demonstrate that additional assumptions are smuggled in when irreversibility is 'derived' from a reversible deterministic theory. A similar type of argument can be made for the supposed derivations of the Second law from a many-particle Schrodinger equation \cite{drossel2015relation}. In both cases, the additional arguments are randomness of initial conditions and of statistical independence of degrees of freedom. The main ideas behind the 'derivations' of the Second Law from classical mechanics are as follows:

(i) A gas is modelled as  a conservative mechanical system of $N \simeq 10^{23}$ small balls with hard-core repulsion, enclosed in a container of volume $V$ with perfectly reflecting walls. Since energy does not change during time evolution, the trajectory of this mechanical system in $6N$-dimensional phase space stays within the $6N-1$-dimensional energy shell in which the initial state is placed. (ii) It is generally assumed, even though this is not proven, that the dynamics of this system is ergodic. This means that a ``typical'' trajectory approches each point of the energy shell with arbitrary precision $\epsilon$ if enough time has passed. Or, equivalently, an initially small and compact phase-space volume of dimension $6N$ (representing an ensemble of very similar initial conditions) will become stretched and folded to the extent that eventually some part of it will be in every small volume of size $\epsilon^{6N-1}$ in the energy shell. (iv) The vast majority of cells of size $\epsilon^{6N-1}$ correspond to macrostates with maximum entropy, which have (among other properties) an even distribution of density over the entire volume. (v) Therefore, if starting in one of the few cells that correspond to a low-entropy initial state, after sufficiently long time the state of the system will 'almost certainly' be in a cell of maximum entropy.

Of course, each of these steps could be discussed in depth, in particular the starting assumption of infinite-precision phase space points and infinite-precision trajectories. I will focus on the last step and accept all preceding ones for the sake of brevity. In this last step, additional assumptions creep in: The last step is based on the assumption that the trajectory taken by the system is a 'typical' trajectory, or, equivalently, that among all the possible initial states that lie within the initial energy-shell volume element, the true initial state is a randomly chosen one. This is an assumption of randomness of the initial state. The laws of classical mechanics would not be violated if all particles of the gas would gather in irregular time intervals in some corner of the container. Among all possible initial states that agree with each other within a desired degree of precision, there would be particular initial states that show such an atypical behavior. We therefore have to introduce the additional assumption that the 'true' initial state is not such an 'atypcial' one that leads to unexpected low-entropy future states. This is somewhoat similar to what one does in electrodynamics when ruling out spatially localized sources of radiation that lie in the future.

Instead of postulating random initial conditions, one could equivalently argue that the initial state is not fixed with infinite precision but only with finite precision, and that randomness comes in as the time evolution of the system proceeds and more and more  bits that were not fixed initially become relevant for fixing the ongoing time evolution. In my view there are many reasons to prefer this latter perspective, and it has been defended most forcefully by Nicolas Gisin \cite{gisin2020real}.  

But there is more to stochasticity than that it is always added to 'deterministic' theories when real-world problems are addressed. Stochasticity depends on context. The next section focuses on this issue.

\section{Stochasticity is context dependent}

We have seen above that whenever the environment of a system is included irreversibility comes in: In mechanical systems, the environment causes friction. In electrodynamics, considering the emission of radiation means that there is an environment into which this radation can be emitted. In quantum mechanics, the interaction of a quantum particle with the rest of the world, which is here represented by a measurment device, causes the irreversible measurement event. Thermodynamics is a very interesting field of science from this point of view as it is a prime example of the environment determining what happens in a system: The environment imposes the temperature and the chemical potential. The environment performs changes on the system (such as pushing a piston or connecting a hot and a cold object) the effect of which is calculated in thermodynamics problems.

It is usually not acknowledged that stochasticity also requires a context. Karl Popper emphasized it in his paper ``The propensity interpretation of probability''\citep{popper1959propensity}, where he argues that probabilities are defined only when the conditions under which the experiment is to be repeated are specified. Probabilities are thus defined relative to  a setup in which the event of interest occurs. Popper justifies the propensity view by the findings of quantum mechanics. Indeed, the ``fundamental'' stochastic event is a quantum event. Its standard example is a measurement. Let us take for illustrative purposes the Stern-Gerlach experiment. It is the simplest quantum measurement as the observable to be measured is a spin-1/2, with only two measurement results, which we denote as $|+\rangle$ and $|-\rangle$. The measurement device measures the spin with respect to direction chosen by the experimenter. Let this direction be the positive $z$ direction. If the incoming particles have not been polarized such that they point in the $z$ direction, both measurement outcomes $|+\rangle$ and $|-\rangle$ are possible, and their probabilities depend on the preparation procedure. Now, the experimenter could have chosen to measure the spin with respect to another direction, for instance the $x$ direction. In this case the possible measurement results would be again  $|+\rangle$ and $|-\rangle$, but now these two results signify an orientation parallel or antiparallel to the direction of the $x$ axis and no longer to the $z$ axis. In general, the probabilities for the two measurement results are also different compared to a measurement in $z$ direction. 

This example demonstrates that the possible events and their probabilities are not an intrinsic property of the quantum particle but the combined effect of the way how the quantum particle has been prepared and of the measurement device. Had the experimenter chosen to measure position or energy instead of spin, the possible results and their probabilities would again have been different. Had the experimenter decided not to perform any measurement, the quantum particle would have remained in the state generated by the preparation procedure.

Now, quantum measurements are just one class of quantum mechanical events by which a quantum particle undergoes an irreversible, stochastic transition and exchanges energy with a macroscopic, finite-temperature object (which is a classical object). There need not be an experimenter, the particle could interact with the rest of the world in a different way. Examples are nuclear fusion reactions in the core of the sun or the emission of photons into space from the warm surface of the earth. Again, the fact that these events happen and their probabilities depend on the context. 

So far, not many models for quantum measurement take this context explicity into account. Many authors, however, speak in a more general sense of quantum contextuality \citep{grangier2018quantum,jaeger2020quantum}. The Copenhagen interpretation of the measuremeent process makes an explicit distinction between classical objects (the measurment setup) and quantum objects, but does not detail how this difference arises. More explicit considerations of how the context is involved in a quantum measurement are advanced by  Ellis \cite{ellis2012} and more recently by Drossel and Ellis \cite{drossel2018contextual}.  

This insight that there is a top-down effect from the environment or the larger spatial scale to the quantum particle leads us to the next section where we introduce the concepts of  contextual emergence and top-down causation. 

\section{Contextual emergence and top-down causation}

Physics is viewed by many people as a reductionistic science where the properties of an object are derived from its parts and their interactions. Consequently, those theories that deal with the most microscopic objects (i.e., particle physics and quantum field theory) are considered to be the most fundamental ones and to describe, at least in principle, everything that occurs on larger scales. But this is a one-sided view that ignores the various ways in which the context or the larger structure influences what happens at a smaller scale. To understand this better, it is useful to think of nature as a hierarchy of objects, with the objects on the lower level being the parts of the objects on the next hierarchical level. For instance, one can build the hierarchy Elementary particle - Atom - Crystal - Earth - Solar system - Galaxy - Universe; or, choosing a hierarchy that involves us humans: Elementary particle - Atom - Molecule - Cell - Organ - Individual - Society.

Now, describing an object in terms of its parts and their interaction is practised successfully by all scientists.
For instance, describing the conducting properties of a metal, physicists build a model of electrons performing collisions with the lattice defects and the lattice vibrations (the so-called phonons) of the metal.
Or, when exploring the cycle of growth and division of a biological cell, scientists describe this process by a network of molecular interactions and reactions. This success of the reductionist procedure often veils the view for the top-down influences that are equally important:
The metal block provides the environment for lattice vibrations to exist, and the temperature of the metal block determines how many phonons of which frequency are present. The lattice structure of the metal determines what the conduction bands of the metal look like and thus strongly affects the way how electrons can propagate in the material. The wider context, such as the connection to a power supply is the prerequisite for conduction happening at all.
Similarly, the growth and division of cells is regulated by mechanical and chemical cues from the embedding tissue to which the cell belongs. Furthermore, the cell depends on a continuous supply of nutrients and building material from its environment. 
For any other system of the hierarchies mentioned above, we can similarly list a variety of top-down influences of the whole on its parts. This topic of top-down causation is discussed extensively by George Ellis in his book ``How can physics underlie the mind'' \citep{ellistopdown}.

All this means that reduction, successful as it is, is only part of the
story. The complement of reduction is emergence: while reduction considers how
the whole can be explained in terms of its parts, the concept of emergence
considers the qualitatively new properties of the whole, which are not
properties of the parts. One distinguishes between different concepts of
emergence. Proponents of weak emergence hold that the emergent properties,
even though they appear surprising and qualitatively different from the
properties of the parts, can ultimately be fully accounted for by the
properties of the parts. For a physicist this means that the theory describing
the parts and their interactions contains at least implicitly all the
phenomena observed in the whole. It is closely associated with a reductionist
view. In contrast, proponents of strong emergence hold that the emergent
properties are explained only partially by the parts and their interactions
and cannot be fully reduced to them. The notion that captures best the type of
strong emergence that occurs in physics is that of contextual emergence. It
emphasizes that the existence of the parts and their properties depend on the
context, the boundary conditions, etc. Until know, strong (or contextual) emergentists are a minority among physicists. The reductionist paradigm prevails despite the many good reasons to question it.

With this understanding of contextual emergence and top-down causation, we will gain a deeper understanding of how indeterminism enables our experience of the passage of time and of being agents.

\section{Indeterminism and top-down causation}
\label{bioexamples}

It is often argued that indeterminism is of no help at justifying free will and agency. The reason given is that stochasticity is completely random in the sense that random events are not influenced by anything else and can therefore not be part of a conscious act or decision - which is not random but follows logical reasoning or is based on values. However, our discussion of the quantum measurement process above has shown that stochastic events are not completely random. On the contrary, the possibility of these events and the nature of these events is set by the context. This is not appreciated very often.
Karl Popper made the interesting suggestion that chance at the lower level is necessary for top-down causation from the higher level \citep{Popper1977-POPTSA}, and he has been criticized for it \citep{sep-properties-emergent}. But I think that he is right, and so do others \citep{del2019physics}. Only when the entities at the lower level are not fully controlled by the microscopic laws can they respond to the higher level.

In biological systems, the influence of the environment on the types of stochastic events and their effects is evident at all levels: Let us take gene expression. The process by which transcription factors bind to the promoter region of a gene and enable transcription has many random elements, as diffusion and binding of transcription factors is subject to thermal noise. However, the context determines which transcription factors are activated at all. For instance, when I experience joy, this triggers the discharge of endorphins, which in turn stimulates the production of further endorphines via gene expression. On a larger timescale, evolution has shaped the transcription factors that we possess and the affinities to the different genes, resulting in the binding and dissociation rates that characterize the stochastic dynamics in our cells.

As another example, consider the brain. The neurons in our brain fire ``randomly''. At the same time, this electric activity reflects the brain activity, for instance while I formulate the text of this paper and type it. There is a wider context that influences the stochastic events in the brain: The past experiences have shaped the connection patterns and the strengths of the synapses of our neural network, and this in turn affects how activity propagates through the network and which neurons or groups of neurons are triggered by which others.  
Although there is randomness in the individual firing events of neurons, and on a more microscopic level in the opening and closing of ion channels, the frequency of firing and the neurons involved and the temporal sequence of which regions become active and which output neurons pass the signal on (for instance to my fingers while I am typing) is not random at all. Somehow it even reflects my creative activity while I develop my arguments for this text. 

The constructive role of randomness in biological systems is discussed in more depth in several review articles \cite{noble2018harnessing,wiesenfeld1995stochastic,tsimring2014noise}. 

\section{Conclusions}

The view of physics that I have defended in this article is very different from the reductionist view that is propagated in many textbooks. The main problem with that view is that theories are applied far beyond their range of applicability. Any physical theory is an approximation and not an exact description of the physical processes of interest. And by no means it is a complete description of everything that is going on in the physical world. I have argued that even though the supposedly fundamental theories are deterministic and time-reversible, these theories are in practice supplemented by irreversible and stochastic features. There is no reason to assume that irreversibility and stochasticity are only apparent. Such a claim is based on ideology and not on evidence. I have argued that irreversibility and stochasticity enter when the influence of the environment on a system is taken into account. This focused the attention on the often-neglected topics of emergence and top-down causation. It is the context that sets the stage for the possible events that can happen in the system, and the context also influences the probabilities of these events.

It follows that our experience of time flowing and us being agents that can
influence what happens is not contradicted by physics. Physics enables
everything and underlies everything, but it does not determine everything. On
the contrary, physics appears to be sufficiently complex to allow even life
and consciousness to function on a material platform \cite{ellistopdown}. This
bears some resemblence to Turing machines in computational science, which are
complex enough to allow for any possible type of calculation.

Top-down influences come in different
shapes. Top-down causation from the wider material context is easier to
understand but is only part of the story. In particular the two biological
examples of Section \ref{bioexamples} illustrate that there is also an
influence from the nonmaterial world of ideas, rationality, goals and desires
on the activity of our genes and neurons. It may well be that physics will
never find an access to describing these influences. The goal of my paper is far more modest, namely to make the case that physics cannot rule out that these influences are present. Physics is not causally closed and does not encompass everything that happens in our world.

To conclude, we should not let naive interpretations of the laws of physics constrain our view of reality. On the contrary, we should trust our most immediate experiences that are the basis for us being capable of thinking and acting, and based on these we should critically examine those simplistic metaphysical claims.

\bibliographystyle{ws-procs961x669}
\bibliography{quantumlit_gesamt,emergence}

\end{document}